\documentclass[prl,aps,twocolumn,superscriptaddress,amsmath,amssymb,floatfix,showpacs,showkeys]{revtex4-1}

\usepackage[colorlinks=true,citecolor=blue,filecolor=blue,linkcolor=blue,urlcolor=blue]{hyperref}

\usepackage{graphicx}           
\usepackage{array}              
\usepackage{placeins}           
\usepackage{amsmath}            
\usepackage{amssymb}            
\usepackage[version=4]{mhchem}  
\usepackage{siunitx}            
\usepackage{xspace}             

\usepackage{verbatim}           
\usepackage{pgf}                
\usepackage[utf8]{inputenc}     
\usepackage{microtype}          

\newcommand{\roughly}{\sim\!}
\newcommand{\rnzero}{\ce{^{220}Rn}\xspace}
\newcommand{\krm}{\ce{^{83$\mathrm{m}$}Kr}\xspace}

\newcommand{\sortofsection}[1]{{\it #1.}---$\!$}    
\newcommand\hrefarxiv[1]{\href{http://arxiv.org/abs/#1}{arXiv:#1}}

\newcommand{\bologna}{\affiliation{Department of Physics and Astronomy, University of Bologna and INFN-Bologna, 40126 Bologna, Italy}}

\newcommand{\chicago}{\affiliation{Department of Physics \& Kavli Institute for Cosmological Physics, University of Chicago, Chicago, IL 60637, USA}}

\newcommand{\coimbra}{\affiliation{LIBPhys, Department of Physics, University of Coimbra, 3004-516 Coimbra, Portugal}}

\newcommand{\columbia}{\affiliation{Physics Department, Columbia University, New York, NY 10027, USA}}

\newcommand{\lngs}{\affiliation{INFN-Laboratori Nazionali del Gran Sasso and Gran Sasso Science Institute, 67100 L'Aquila, Italy}}

\newcommand{\mainz}{\affiliation{Institut f\"ur Physik \& Exzellenzcluster PRISMA, Johannes Gutenberg-Universit\"at Mainz, 55099 Mainz, Germany}}

\newcommand{\heidelberg}{\affiliation{Max-Planck-Institut f\"ur Kernphysik, 69117 Heidelberg, Germany}}

\newcommand{\munster}{\affiliation{Institut f\"ur Kernphysik, Westf\"alische Wilhelms-Universit\"at M\"unster, 48149 M\"unster, Germany}}

\newcommand{\nikhef}{\affiliation{Nikhef and the University of Amsterdam, Science Park, 1098XG Amsterdam, Netherlands}}

\newcommand{\nyuad}{\affiliation{New York University Abu Dhabi, Abu Dhabi, United Arab Emirates}}

\newcommand{\purdue}{\affiliation{Department of Physics and Astronomy, Purdue University, West Lafayette, IN 47907, USA}}

\newcommand{\rpi}{\affiliation{Department of Physics, Applied Physics and Astronomy, Rensselaer Polytechnic Institute, Troy, NY 12180, USA}}

\newcommand{\rice}{\affiliation{Department of Physics and Astronomy, Rice University, Houston, TX 77005, USA}}

\newcommand{\stockholm}{\affiliation{Oskar Klein Centre, Department of Physics, Stockholm University, AlbaNova, Stockholm SE-10691, Sweden}}

\newcommand{\subatech}{\affiliation{SUBATECH, IMT Atlantique, CNRS/IN2P3, Universit\'e de Nantes, Nantes 44307, France}}

\newcommand{\torino}{\affiliation{INAF-Astrophysical Observatory of Torino, Department of Physics, University  of  Torino  and  INFN-Torino,  10125  Torino,  Italy}}

\newcommand{\ucla}{\affiliation{Physics \& Astronomy Department, University of California, Los Angeles, CA 90095, USA}}

\newcommand{\ucsd}{\affiliation{Department of Physics, University of California, San Diego, CA 92093, USA}}

\newcommand{\wis}{\affiliation{Department of Particle Physics and Astrophysics, Weizmann Institute of Science, Rehovot 7610001, Israel}}

\newcommand{\zurich}{\affiliation{Physik-Institut, University of Zurich, 8057  Zurich, Switzerland}}

\newcommand{\paris}{\affiliation{LPNHE, Universit\'{e} Pierre et Marie Curie, Universit\'{e} Paris Diderot, CNRS/IN2P3, Paris 75252, France}}

\newcommand{\freiburg}{\affiliation{Physikalisches Institut, Universit\"at Freiburg, 79104 Freiburg, Germany}}

\newcommand{\lal}{\affiliation{LAL, Universit\'e Paris-Sud, CNRS/IN2P3, Universit\'e Paris-Saclay, F-91405 Orsay, France}}

\newcommand{\naples}{\affiliation{Department of Physics ``Ettore Pancini'', University of Napoli and INFN-Napoli, 80126 Napoli, Italy}} 

\newcommand{\nagoya}{\affiliation{Kobayashi-Maskawa Institute for the Origin of Particles and the Universe, Nagoya University, Furo-cho, Chikusa-ku, Nagoya, Aichi 464-8602, Japan}}

\begin{document}

\title{Light Dark Matter Search with Ionization Signals in XENON1T}
\author{E.~Aprile}\columbia
\author{J.~Aalbers}\email[]{jelle.aalbers@fysik.su.se}\stockholm
\author{F.~Agostini}\bologna
\author{M.~Alfonsi}\mainz
\author{L.~Althueser}\munster
\author{F.~D.~Amaro}\coimbra
\author{V.~C.~Antochi}\stockholm
\author{E.~Angelino}\torino
\author{F.~Arneodo}\nyuad
\author{D.~Barge}\stockholm
% ...

\collaboration{XENON Collaboration}
\email[]{xenon@lngs.infn.it}
\noaffiliation

\date{\today}

\begin{abstract}
We report constraints on light dark matter (DM) models using ionization signals in the XENON1T experiment. 
We mitigate backgrounds with strong event selections, rather than requiring a scintillation signal, leaving an effective exposure of $(22 \pm 3)$ tonne-days. 
Above $\roughly\SI{0.4}{keV_{ee}}$, we observe $<1 \, \text{event}/(\text{tonne} \times \text{day} \times \text{keV}_\text{ee})$, which is more than one thousand times lower than in similar searches with other detectors. 
Despite observing a higher rate at lower energies, no DM or CEvNS detection may be claimed because we cannot model all of our backgrounds.
We thus exclude new regions in the parameter spaces for DM-nucleus scattering for DM masses 
$m_\chi$ within $3-\SI{6}{GeV/c^2}$, 
DM-electron scattering for $m_\chi > \SI{30}{MeV/c^2}$,
and absorption of dark photons and axion-like particles for $m_\chi$ within $0.186 - \SI{1}{keV/c^2}$.

\end{abstract}

\pacs{
    95.35.+d,  
    14.80.Ly,  
    29.40.-n,  
    95.55.Vj   
}


\maketitle


\sortofsection{Introduction} 
Substantial cosmological and astrophysical observations show that much of the Universe's mass consists of dark matter (DM)~\cite{planck, wimp_hooper}, and experiments aim to detect hypothetical DM particles and identify their nature~\cite{wimp_review, directdetection_review}. The XENON1T experiment recently set the world's most stringent limits on DM-nucleus scattering for DM masses $m_\chi \geq \SI{6}{GeV/c^2}$~\cite{sr1prl, x1t_sd}. This paper reanalyzes XENON1T's data to constrain lighter DM.

XENON1T~\cite{1t_instrument} is a dual-phase time projection chamber (TPC) housed at the INFN Laboratori Nazionali del Gran Sasso. The active volume contains 2 tonnes of liquid xenon (LXe) and is bounded by a grounded gate electrode at the top ($z=0$) and a cathode at the bottom ($z=\SI{-97}{cm}$). Charged particles recoiling in LXe produce photons (scintillation) and electrons (ionization). XENON1T promptly detects the photons as the `S1' signal with 248 3-inch photomultiplier tubes (PMTs) positioned above and below the LXe target~\cite{1t_pmts, 1t_pmts2}. Electric fields drift the electrons upward and extract them into gaseous xenon, where electroluminescence produces a secondary scintillation `S2' signal. In most analyses, the ratio between S1 and S2 differentiates electronic recoils (ERs), caused by $\beta$ particles and $\gamma$ rays, from nuclear recoils (NRs), caused by neutrons or some DM particles. The interaction position, reconstructed from the S2 light pattern and the time difference between S1 and S2, discriminates DM candidates from most external radioactive backgrounds.
XENON1T is shielded by a \SI{3600}{m} water-equivalent rock overburden, an active water Cherenkov muon veto~\cite{1t_mv}, and 1.2 tonnes of LXe surrounding the TPC.

Dual-phase LXe TPCs are most sensitive to DM with masses $m_\chi \gtrsim \SI{6}{GeV/c^2}$, as lighter DM cannot transfer enough energy ($\roughly \SI{3.5}{keV}$) to xenon nuclei to yield detectable S1s at a sufficient rate to be useful in DM experiments. 
S2s, however, are detectable in LXe at energies as low as 0.7~keV for nuclear recoils and 0.186~keV for electronic recoils~\cite{lux_dd, lux_lower}. Here, we reanalyze XENON1T's data without requiring an S1 -- an `S2-only analysis'.
As in previous S2-only analyses~\cite{x10_s2only, x100_s2only}, substantial backgrounds preclude detection claims. However, we use strong event selections to reduce these backgrounds considerably and subtract some known background components.

\begin{figure}
    \centering
    \includegraphics[width=0.5\textwidth]{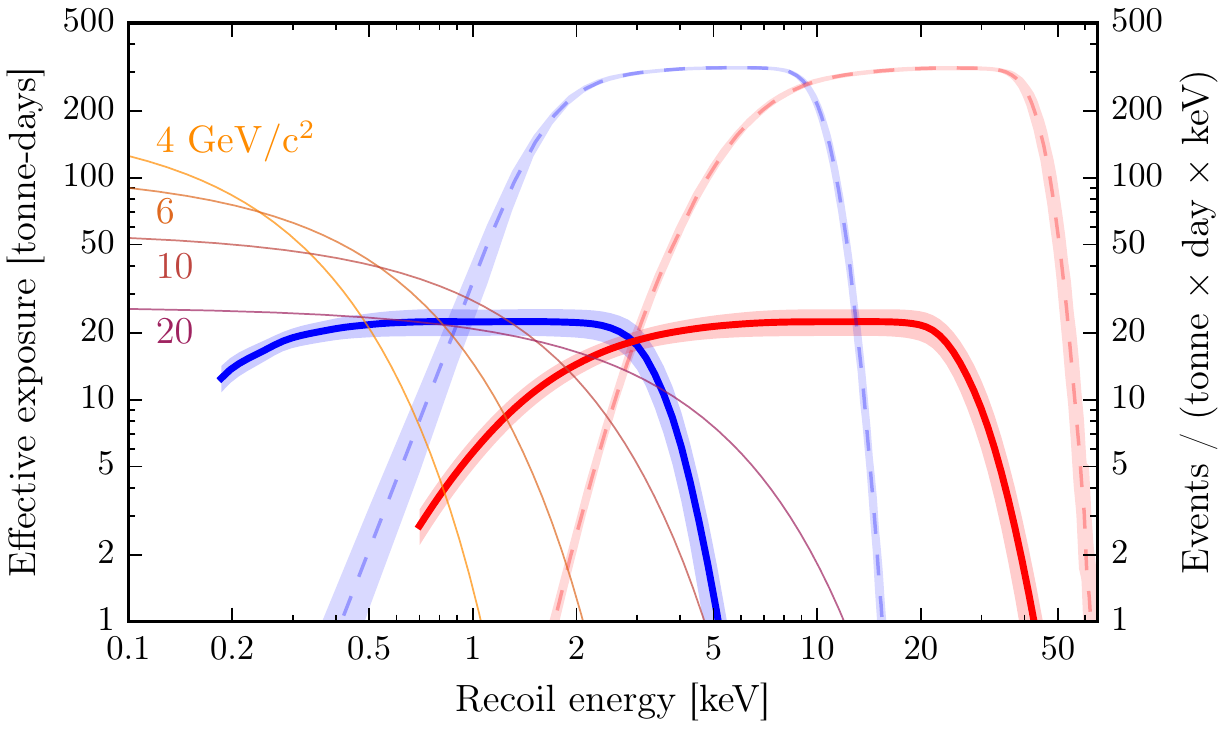}
    \caption{Effective remaining exposure after event selections for NR (red) and ER (blue) signals of different energies, for S2 $\in [150, 3000]$ PE, on the left y-axis. Dashed lines show the same for XENON1T's main analysis \cite{sr1prl}, and shaded bands show $\pm 1 \sigma$ systematic uncertainties. Thin lines show the expected differential event rate for 4, 6, 10, and \SI{20}{GeV/c^2} spin-independent (SI) DM-nucleus scattering with $\sigma = 10^{-43}\si{cm^{2}}$, under the nominal signal model, on the right y-axis.}
    \label{fig:total_eff}
\end{figure}

\sortofsection{Data selection}
We use the main science run (SR1) of XENON1T~\cite{sr1prl, x1t_sd} with a livetime of 258.2~days, after excluding time when the data acquisition was insensitive, the muon veto fired, or a PMT showed excessive pulse rates \cite{1t_instrument}. 
Ref.~\cite{sr1prl} derived a $\roughly4\%$ shorter livetime because it excluded time just after high-energy events. Backgrounds from these periods are mitigated by other methods here.

We used 30\% of SR1 events as training data, distributed homogeneously in time, to determine event selections and, for each dark matter model and mass, a region of interest (ROI) in terms of the integrated S2 charge (`S2 size'). Limits on DM parameters are computed using only the remaining 70\% (the `search data', 180.7 days), which was not examined with the strong event selections used here until the analysis was fixed. We chose selections to remove large identifiable background populations, but retain the central part of the DM signal model in different observable dimensions. We use a single set of selections; only the S2 ROIs vary for different DM models and masses. 

\begin{figure}
    \centering
    \includegraphics[width=0.5\textwidth]{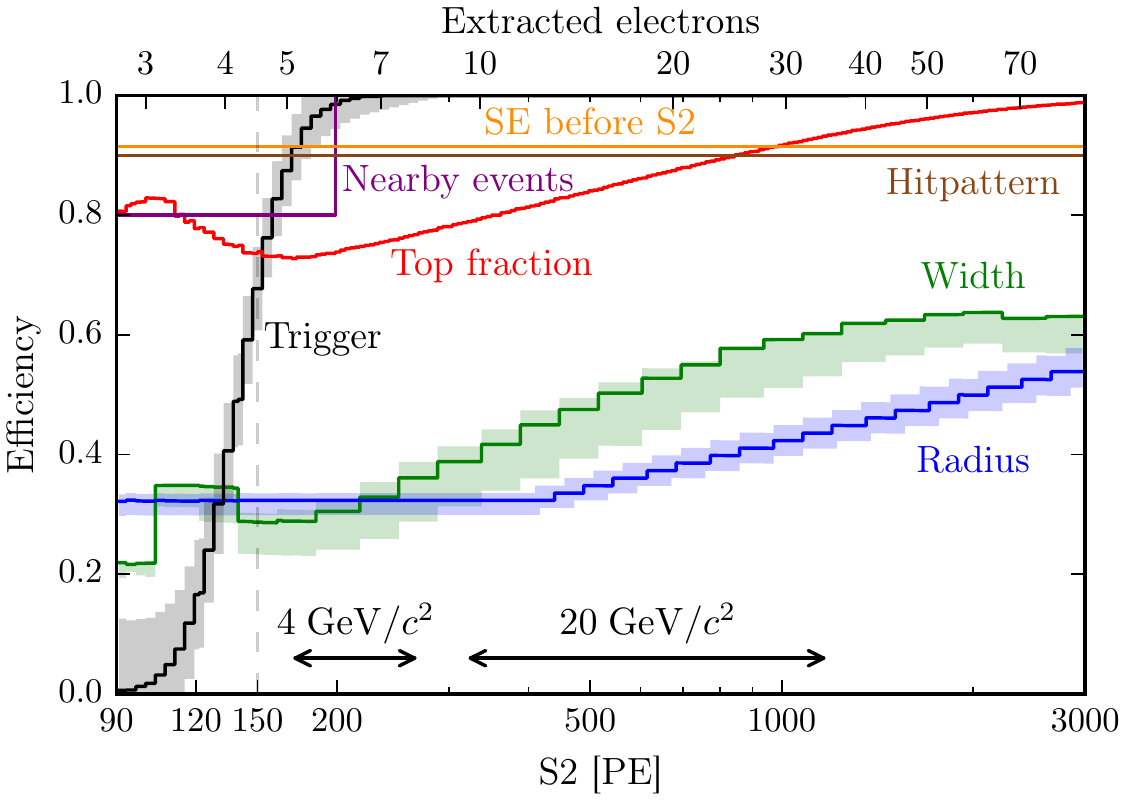}
    \caption{Efficiencies (fraction of signal events passed) of the most impactful event selections versus S2 size for $z \in [-30, -10]$~cm, applicable if a given selection is applied last. Solid lines correspond to the nominal detector response model, bands to $\pm 1 \sigma$ variations of model parameters. The arrows show the S2 ROIs for the \SI{4}{GeV/c^2} and \SI{20}{GeV/c^2} spin-independent NR DM analyses. Events below \SI{150}{PE} are not used but shown for completeness. The top horizontal axis shows the approximate number of extracted electrons corresponding to each S2 size. The combined efficiency of the selections not shown here is $\roughly93\%$.}
    \label{fig:effs}
\end{figure}

We determine the efficiency of our selections in the 2D space of (uncorrected) S2 signal size and interaction depth $z$ -- since our signal models vary strongly along both dimensions -- using calibration data and simulated waveforms~\cite{ap1}.
Figure \ref{fig:total_eff} shows the effective remaining search data exposure after selections. Figure~\ref{fig:effs} shows the efficiencies of the most impactful cuts in the most important $z$ range for light DM.
XENON1T's trigger efficiency, shown in black in Figure \ref{fig:effs}, is determined as in~\cite{daqpaper}.
Events with S2 below 150 photoelectrons (PE), i.e.~$\roughly4.5$ extracted electrons, are not used, but they are shown for completeness.
Previous XENON1T analyses~\cite{sr0prl, sr1prl, x1t_sd, wimppion} applied a similar threshold of \SI{200}{PE}; the S1 requirement, not the S2 threshold, limited their light DM sensitivity.

\begin{figure}
    \centering
    \includegraphics[width=0.5\textwidth]{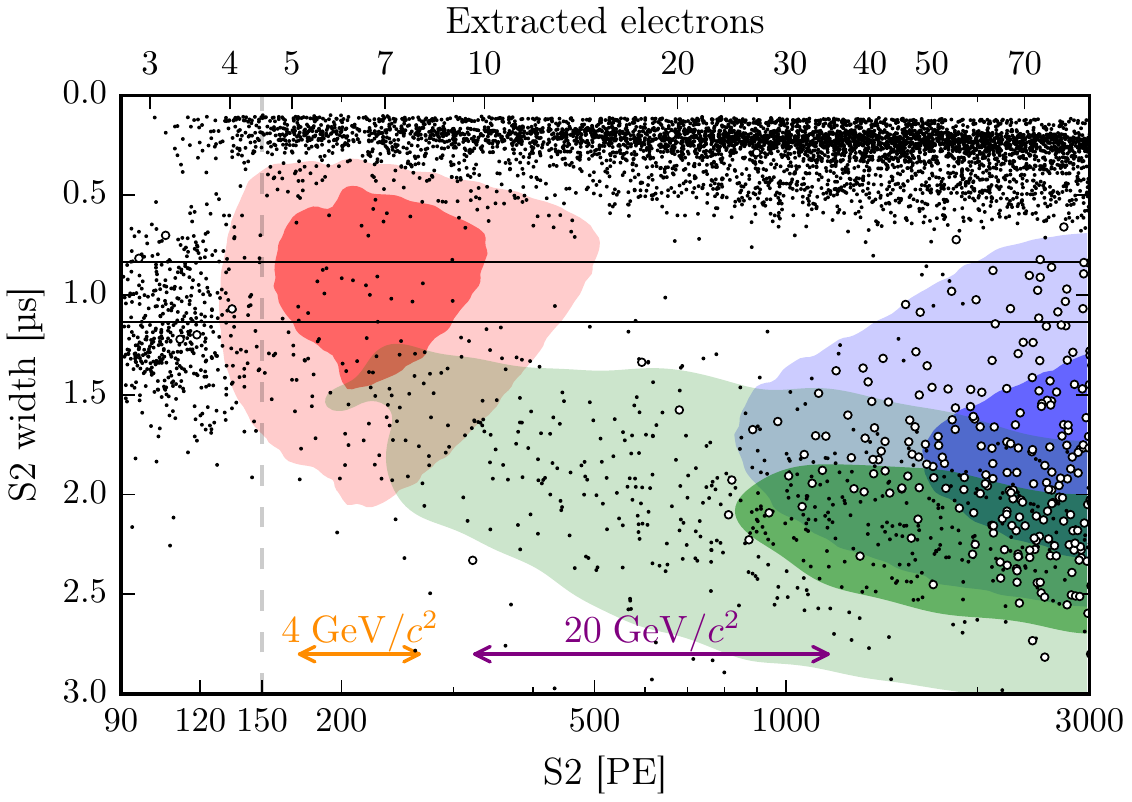}
    \caption{Observed events in the search data. Events between the horizontal black lines pass all cuts, the others fail only the cut on S2 width. Open circles demark events with S1s, dots events without. The colored regions contain 50\% (faint: 90\%) of expected events from the three background components described in the text: flat-spectrum ER (blue), CEvNS (red), and cathode events (green). The arrows denote two S2 ROIs and the dashed line the $\mathrm{S2}$ threshold, as in Fig.~\ref{fig:effs}.}
    \label{fig:monster}
\end{figure}

Without S1s, the event depth $z$ cannot be accurately estimated. However, the S2 waveform width in time is correlated with $z$, due to diffusion of the electrons during drift~\cite{sorensen_diffusion}. Rather than estimating and constraining $z$, we remove events with S2 width outside $[835, 1135]~\si{ns}$, as shown in Figure~\ref{fig:monster}. This window has a high expected signal rate, but showed few events in the training data, indicating a low background. The width cut mitigates backgrounds with atypically wide S2s, consistent with $\beta$ decays occurring on the cathode wires. The field geometry there causes charge loss, which in turn causes the cathode events to have unusually small S2s for their energy. Many have detectable S1s and are called `S1-tagged cathode events'; these are easily removed by another cut (described below). The width cut also mitigates backgrounds with atypically narrow S2s, which could similarly result from decays on the electrodes at the top of the TPC.

The width cut efficiency, calculated with simulated S2 waveforms, is shown in green in Figure~\ref{fig:effs}. The simulated waveform's median widths agree to within $\roughly\SI{50}{ns}$ with those observed in deuterium-deuterium plasma fusion neutron generator calibration data~\cite{ng}, as detailed in the supplement. The cut efficiency is highest in the $z \in [-30, -10]~\si{cm}$ range presented in Figure~\ref{fig:effs}, where the expected DM signals are most distinguishable from backgrounds. We include a $\pm \SI{50}{ns}$ systematic uncertainty on the S2 width, creating a $12\%$ uncertainty in the expected \SI{4}{GeV/c^2} spin-independent NR DM event rate.

We remove events reconstructed at high radii $R$. The threshold is $R^2 = \SI{700}{{cm}^2}$ for $\mathrm{S2} \leq \SI{400}{PE}$ and then rises linearly with $\log(\mathrm{S2})$ to \SI{1150}{{cm}^2} at \SI{3000}{PE}.
As described in~\cite{ap2}, events on the TPC wall ($R=\SI{47.9}{cm}$) have unusually small S2s because electrons are lost on the TPC walls. They can be mis-reconstructed inward due to the increased position reconstruction uncertainty for small S2s (\SI{1.8}{cm} at $\mathrm{S2} = \SI{200}{PE}$) and, more importantly, the inhomogeneous drift field. The latter cannot be mitigated as in~\cite{ap2} because S1s are needed to estimate $z$ reliably.
The efficiency of the radial cut, shown in blue in Figure~\ref{fig:effs}, is estimated with \krm calibration data.
We introduce an uncertainty to bracket variations in the estimate from \krm datasets at different times and in a simple geometric calculation. This introduces an $8\%$ uncertainty in the \SI{4}{GeV/c^2} spin-independent NR DM rate.

On average, $\roughly63\%$ of S2 light is seen by the top PMT array, with a $\roughly3\%$ position-dependent variation for which we correct. We remove events in which this fraction is $>66\%$, indicative of events produced in the gaseous xenon phase above the usual secondary scintillation region (`gas events'). The efficiency, shown in red in Figure~\ref{fig:effs}, is calculated from binomial fluctuations in photon detection and a small intrinsic spread measured at high S2. We verified the resulting efficiency is conservative using neutron generator data. The efficiency rises below $\roughly\SI{170}{PE}$ as the trigger preselects S2s to which many PMTs contribute, which is rarer for S2s seen mostly by the top array.

Pile-up of randomly emitted single-electron (SE) signals can be misidentified as S2s from real events. We employ three cuts against this background, without which the population most prominent at $\lesssim\SI{150}{PE}$ in Figure~\ref{fig:monster} would be $\roughly50\times$ larger. First, we remove events whose S2 hitpattern on the top array is inconsistent with that of single scatters, as determined by a likelihood test \cite{ap1}. This cut has a 90\% efficiency, shown in brown in Figure~\ref{fig:effs}, as measured with neutron generator data and S1-tagged cathode events. This cut also removes some unresolved double scatter events, e.g., from radiogenic neutrons. 
Second, we exclude events with one or more S2 or single-electron signals up to $\roughly\SI{1}{ms}$ before the largest S2, with 91.5\% efficiency, as measured with high-energy background events and shown in orange in Figure~\ref{fig:effs}. This cut also suppresses gas events, whose S1s are broader than those of events in the liquid and therefore often misidentified as S2s.
Third, as high-energy events cause a temporary and localized enhancement in single-electron emission~\cite{sorensen_delayed}, we utilize a combined p-value cut \cite{stouffer} against events close in time or reconstructed position to recent high-energy events, with 80\% efficiency, as determined with S1-tagged cathode events and shown in purple in Figure~\ref{fig:effs}. This last cut only helps against the single-electron pileup background, so we apply it only for S2 $< \SI{200}{PE}$. 

We exclude events in which the S2 waveform is distorted by a merged S1, with $\roughly95\%$ efficiency, as determined with \rnzero~\cite{rnsource} and neutron generator data.
To remove double scatters, we apply the same cut to events with substantial secondary S2s as in~\cite{sr1prl, ap1}, with $99.5\%$ efficiency. 

Finally, we apply two cuts specifically to events with S1s. Events whose drift time indicates a $z$ outside $[-95, -7]$~cm are removed, to exclude events high in the detector and S1-tagged cathode events. We assume no signal or background events are produced outside this $z$ region. Our assumption is conservative because this is a limit-only analysis. We also remove events with a very large S1 ($> \SI{200}{PE}$), with negligible efficiency loss.

\sortofsection{Detector response}
We compute XENON1T's response to ERs and NRs in the same two-dimensional (S2, $z$) space used for the efficiencies and project the model after applying efficiencies onto S2 for comparison with data.
We use the best-fit detector response model from~\cite{ap2}, but we assume in our signal and background models that NRs below \SI{0.7}{keV} and ERs below \SI{186}{eV} ($\roughly 12$ produced electrons) are undetectable, as the LXe charge yield $Q_y$ has never been measured below these energies. Even without these cutoffs, the low-energy $Q_y$ from~\cite{ap2} is lower than that favored by other LXe measurements~\cite{lux_dd,lux_lower} and models~\cite{nest_v2}. Thus, our results should be considered conservative.

\begin{figure}
    \centering
    \includegraphics[width=0.5\textwidth]{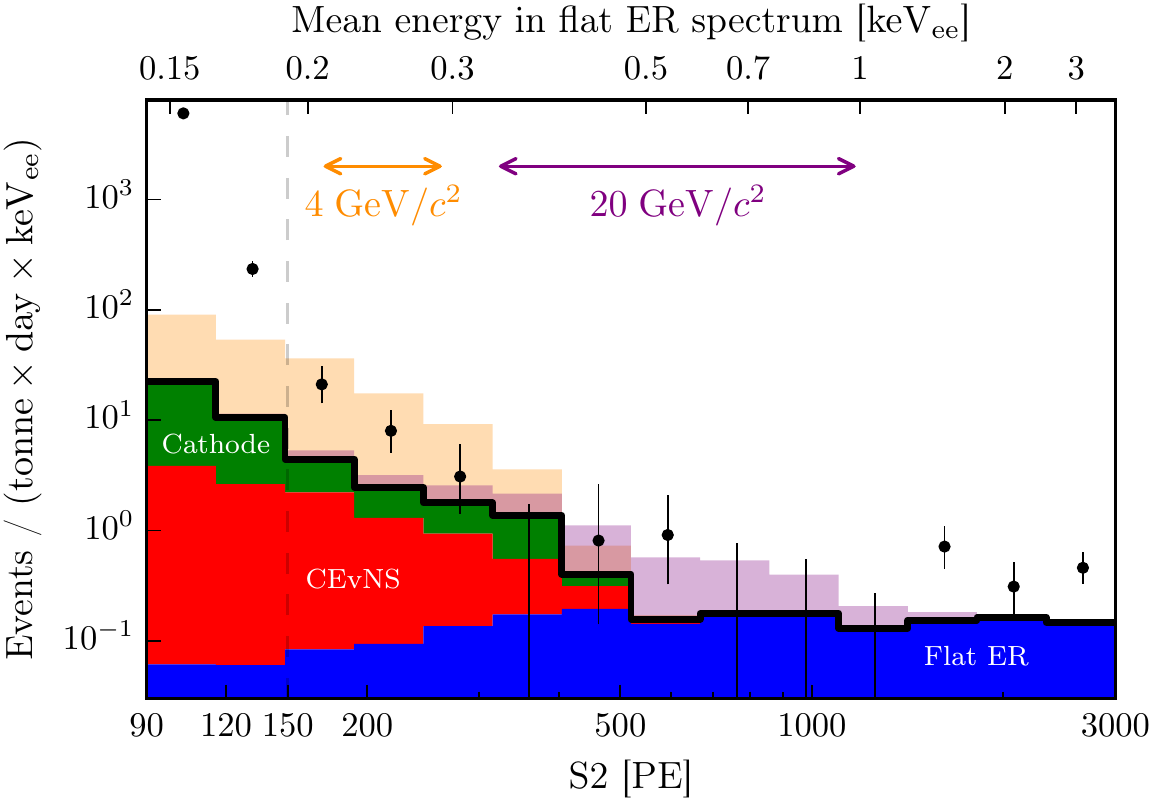}
    \caption{Distribution of events that pass all cuts (black dots); error bars show statistical uncertainties ($1 \sigma$ Poisson). The thick black line shows the predetermined summed background model, below which its three components are indicated, with colors as in Fig.~\ref{fig:monster}. The lightly shaded orange (purple) histogram, stacked on the total background, shows the signal model for \SI{4}{GeV/c^2} (\SI{20}{GeV/c^2}) SI DM models excluded at exactly 90\% confidence level. The arrows show the ROIs for these analyses, and the dashed line the $\mathrm{S2}$ threshold, as in Figures~\ref{fig:effs}-\ref{fig:monster}. All rates are shown relative to the effective remaining exposure after selections. The top x-axis shows the mean expected energy of events after cuts for a flat ER spectrum if there were no $Q_y$ cutoff.}
    \label{fig:stackhist}
\end{figure}

While a complete model of backgrounds in the S2-only channel is unavailable, we can quantify three components of the background, illustrated in Figures~\ref{fig:monster} and \ref{fig:stackhist}. First, 
the ER background from high Q-value $\beta$ decays, primarily \ce{^{214}{Pb}} ($Q = \SI{1.02}{MeV}$)~\cite{ap2}, is flat in our energy range of interest. We use a rate of \SI{0.142}{events/(tonne \times day \times keV)}, a conservative lower bound derived from $<\SI{210}{keV}$ data.
Second, coherent nuclear scattering of \ce{^8B} solar neutrinos (CEvNS), shown in red in Figure~\ref{fig:monster}, should produce a background nearly identical to a \SI{6}{GeV/c^2}, \SI{4e-45}{cm^2} spin-independent (SI) NR DM signal~\cite{cevns, cevns_2}. We expect $2.0 \pm 0.3$ CEvNS events inside the \SI{6}{GeV/c^2} SI NR ROI. 
Third, we see events from $\beta$ decays on the cathode wires. Sufficiently low-energy cathode events lack S1s. We derive a lower bound on this background using the ratio of events with and without S1s measured in a high-S2, high width control region where cathode events are dominant. This procedure is detailed in the supplement.

Figure \ref{fig:stackhist} compares the observed events to our nominal signal and background models. For $\mathrm{S2} \gtrsim \SI{300}{PE}$ ($\roughly\SI{0.3}{keV_{ee}}$), we observe rates well below $1/(\text{tonne} \times \text{day} \times \text{keV}_\text{ee})$, more than one thousand times lower than previous S2-only analyses~\cite{ds_s2only, x100_s2only}. Below \SI{150}{PE}, the rate rises quickly, likely due to unmodeled backgrounds.

\sortofsection{DM models}
We constrain several DM models, using~\cite{wimprates} to compute the energy spectra. First, we consider spin-independent (SI) and spin-dependent (SD) DM-nucleus scattering with the same astrophysical ($v_0$, $v_\text{esc}$, etc...) and particle physics models (form factors, structure functions) as~\cite{sr1prl, x1t_sd}. For SD scattering, we consider the neutron-only (to first order) coupling specifically. If the DM-matter interaction is mediated by a (scalar) particle of mass $m_\phi$, the differential rate has a factor ${m_\phi}^4 / ({m_\phi}^2 + q^2 / c^2)^2$, with $q = \sqrt{2 m_N E_R}$ the momentum transfer, $E_R$ the recoil energy, and $m_N$ the nuclear mass~\cite{lightmediator, lightmediator2, pandax_lm}. Usually, this factor is considered to be $\roughly 1$, corresponding to $m_\phi \gtrsim 100~\si{MeV/c^2}$. We also consider the SI light-mediator (SI-LM) limit, $m_\phi \ll q/c \approx 10^{-3} m_\chi$ (for $m_\chi \ll m_N$), in which the differential event rate for DM-nucleus scattering scales with $m_\phi^4$.

\begin{figure*}
\centering
\begin{tabular}{ >{\centering\arraybackslash}m{8.6cm}  >{\centering\arraybackslash}m{8.7cm}  }
    \includegraphics[width=8.6cm]{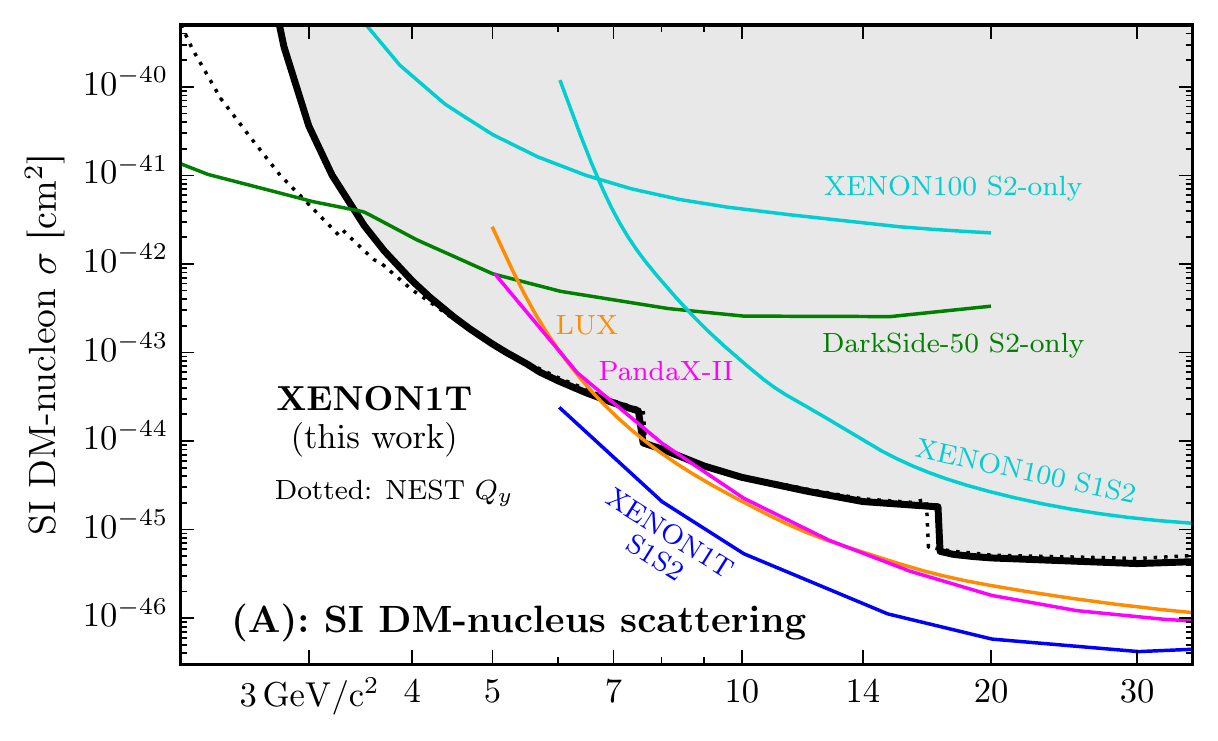} & 
        \includegraphics[width=8.6cm]{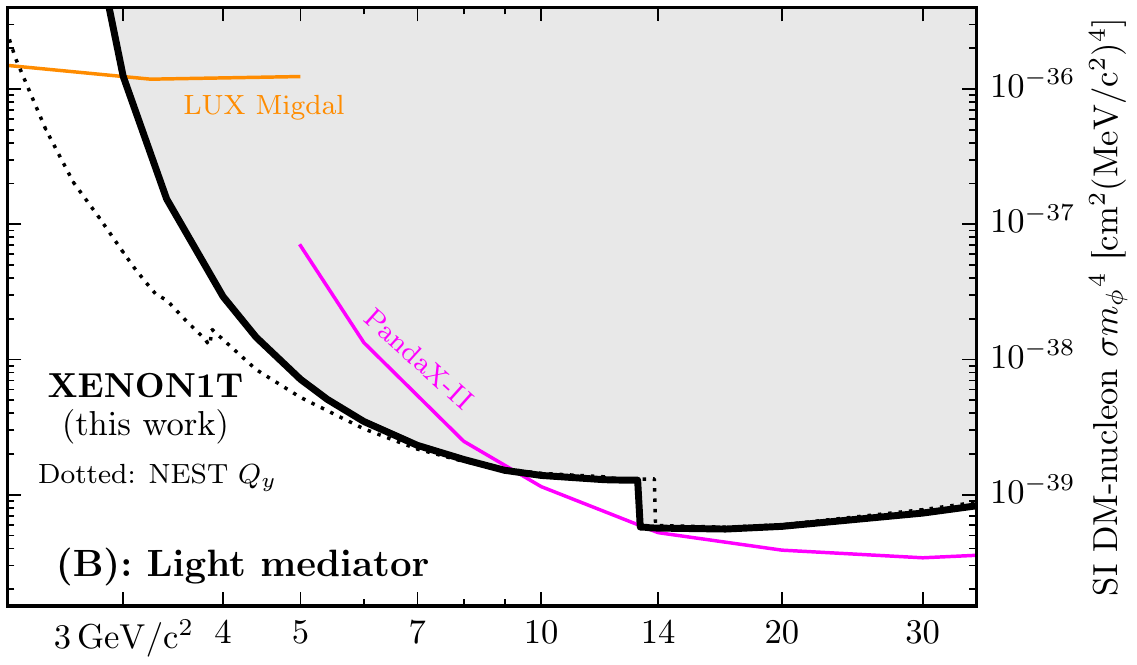} \\
    \includegraphics[width=8.6cm]{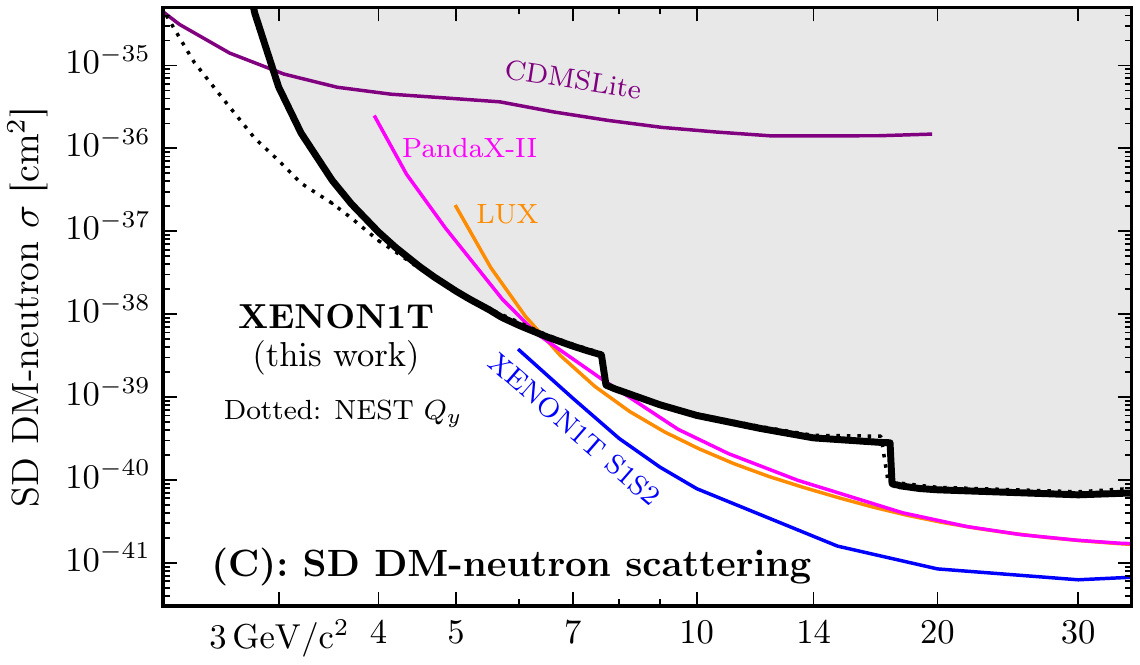} & 
        \includegraphics[width=8.6cm]{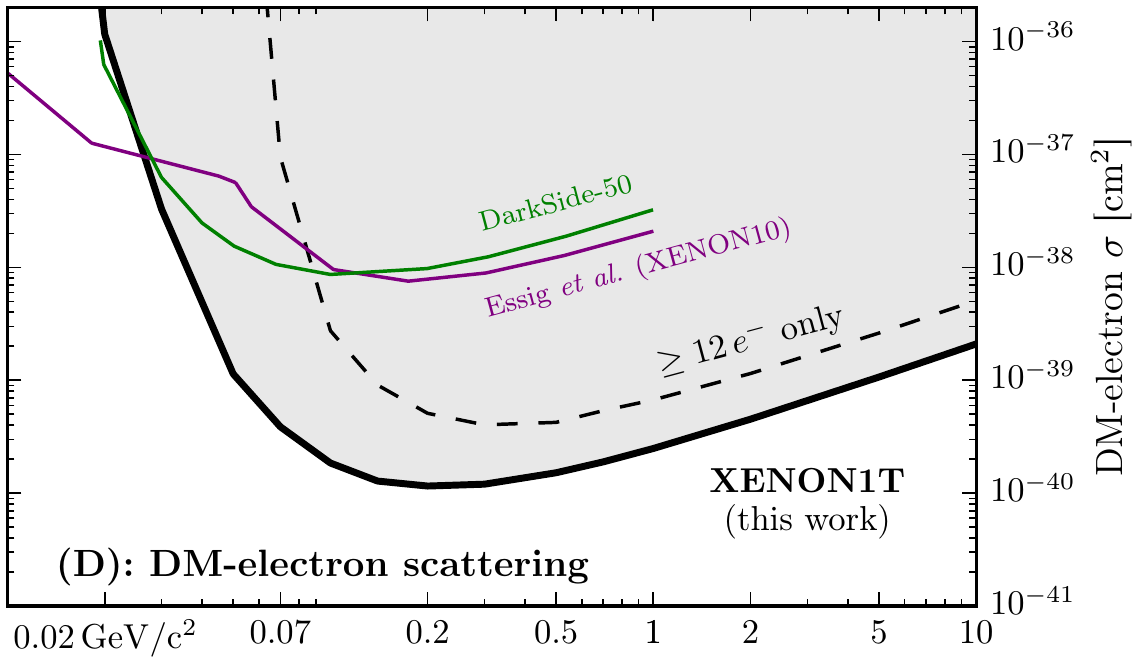} \\
    \includegraphics[width=8.6cm]{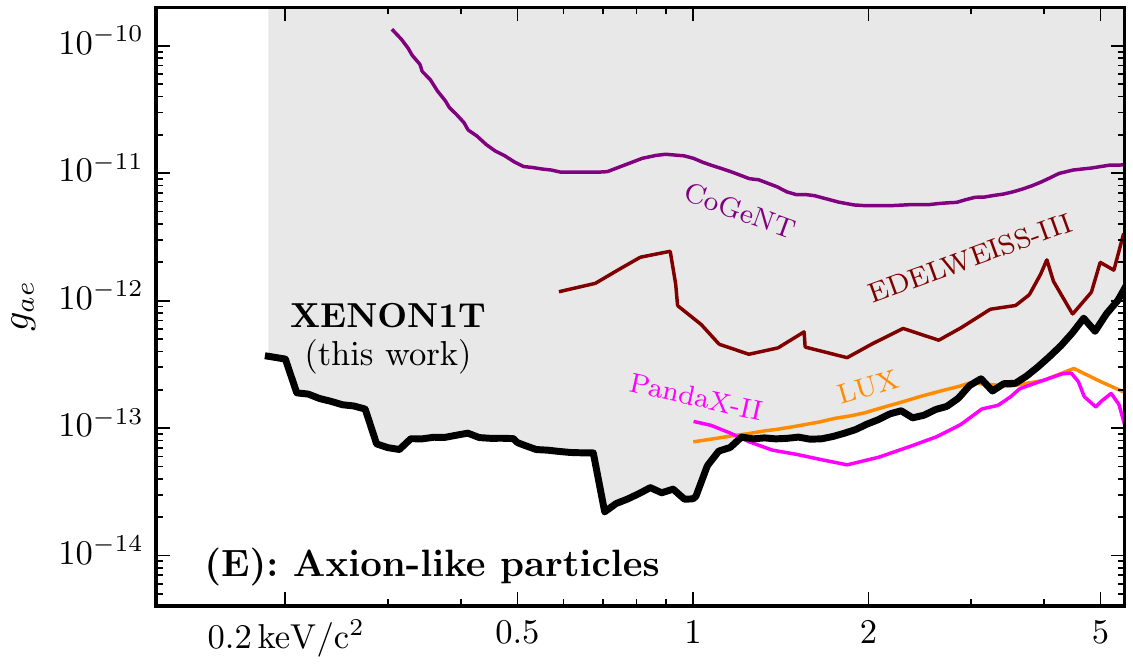} & 
        \includegraphics[width=8.6cm]{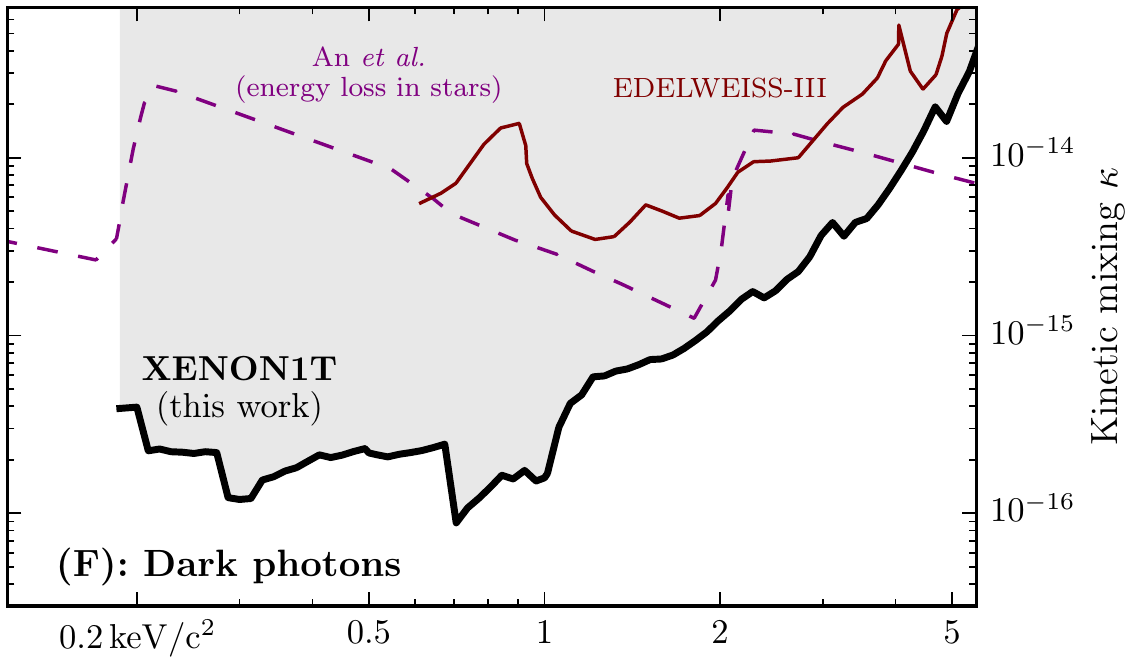}
\end{tabular}
\caption{The 90\% confidence level upper limits (black lines with gray shading above) on DM-matter scattering for the models discussed in the text, with the dark matter mass $m_\chi$ on the horizontal axes. We show other results from XENON1T in blue~\cite{sr1prl, x1t_sd}, LUX in orange~\cite{lux_si, lux_migdal, lux_sd, lux_axions}, PandaX-II in magenta~\cite{pandax_main, pandax_lm, pandax_axions}, DarkSide-50 in green~\cite{ds_s2only, dme_ds, ds_main}, XENON100 in turquoise~\cite{x100_rc, x100_s2only},  EDELWEISS-III~\cite{edelweiss_alp} in maroon, and other constraints~\cite{cogent_alp, cdmslite_sd, dme_x100, stellar_dph} in purple. Dotted lines in panels A-C show our limits when assuming the $Q_y$ from NEST v2.0.1 \cite{nest_v201} cut off below \SI{0.3}{keV}. The dashed line in panel D shows the limit without considering signals with $< 12$ produced electrons; the solid line can be compared to the constraints from~\cite{dme_x100, dme_ds} shown in the same panel, the dashed line to our results on other DM models, which use the $Q_y$ cutoffs described in the text. The limits jump at \SI{17.5}{GeV/c^2} in panel A (and similarly elsewhere) because the observed count changes from 10 to 3 events in the ROIs left and right of the jump, respectively.}
\label{fig:limits}
\end{figure*}

Second, light DM could be detected from its scattering off bound electrons. We follow~\cite{dme_x100} to calculate the DM-electron scattering rates, using the ionization form factors from~\cite{tty_site}, the detector response model as above (from~\cite{ap2}), and dark matter form factor $1$. 
Relativistic calculations~\cite{dme_relativistic} predict $2-10\times$ larger rates (for $\geq 5$ produced electrons), and thus our results should be considered conservative.
As previous DM-electron scattering results~\cite{dme_x10, dme_x100, dme_ds} did not use a $Q_y$ cutoff, we derive constraints with and without signals below 12 produced electrons (equivalent to our $Q_y$ cutoff) to ease comparison.

Third, bosonic DM candidates, such as dark photons and axion-like particles (ALPs), can be absorbed by xenon atoms, analogous to photons in the photoelectric effect. The result is a monoenergetic ER signal at $E_\chi = m_\chi c^2$, with rates of
\begin{equation*}
\begin{bmatrix}
    \SI{4e23}{keV} \cdot \kappa^2 / E_\chi \\
    \SI{1.3e19}{keV^{-1}} \cdot {g_\text{ae}}^2 E_\chi
\end{bmatrix}
 \frac{\sigma_\text{pe}}{A} \si{{kg}^{-1} {day}^{-1}},
\end{equation*}
where the top row corresponds to dark photons~\cite{dph_theory} and the bottom to ALPs~\cite{alp_theory}. Here $\sigma_\text{pe}$ is xenon's photoelectric cross-section at $E_\chi$ in barn, $A$ xenon's mean atomic mass number, $\kappa$ the dark photon-photon kinetic mixing parameter, and $g_\text{ae}$ the axioelectric coupling constant. This process allows us to constrain \si{keV}-scale DM candidates.

\sortofsection{Inference and Results}
We constrain these DM models based on the number of events in pre-determined S2 ROIs, which vary for each model and mass. The ROIs are optimized to give stringent limits on the training data, while requiring that the lower (upper) bound is between the 5th and 60th (40th and 95th) percentile of the signal distribution in $\mathrm{S2} \in [90, 3000]~\si{PE}$ after selections, and never below $\SI{150}{PE}$. 
These constraints, the event selections, and the background models were set before examining the search data.
Due to the finite training data, the ROI bounds are non-smooth functions of DM mass.

We compute an aggregate uncertainty on the signal and background expectations in the ROIs, including the $\roughly5\%$ uncertainty on electron lifetime and $\roughly2.5\%$ uncertainty on the S2 gain $g_2$, besides the (more impactful) uncertainties on the efficiencies mentioned above. We then compute 90\% confidence level upper limits using the standard Poisson method~\cite{pdg_stats}. To ensure the limits are conservative (statistically over-cover), we use the 10th percentile signal and background expectations -- i.e.~the nominal value minus $\roughly1.28 \times$ the aggregate uncertainty -- and never exclude signals with $<2.3$ expected events.

The resulting DM upper limits are shown in Figure~\ref{fig:limits}. We exclude new regions in the parameter spaces for all DM models shown. ER models (panels D-F in Figure~\ref{fig:limits}) benefit most from an S2-only analysis, as ERs produce smaller S1s than NRs at the same S2 size. 
Our constraints on \SI{6}{GeV/c^2} SI NR DM, and therefore $\ce{^8B}$ CEvNS, are weaker than in \cite{sr1prl} due to the 16 observed events in the \SI{6}{GeV/c^2} ROI and an under-fluctuation of the background in \cite{sr1prl}. Dotted lines in figure \ref{fig:limits} $\!\!$A-C show our constraints given the NR $Q_y$ from NEST v2.0.1 \cite{nest_v201} cut off below \SI{0.3}{keV}, which accommodates a measurement \cite{livermore_qy} released shortly after our analysis was completed.
Future S2-only studies can improve on these results using next-generation detectors such as XENONnT and LZ~\cite{lz_tdr}, lower-energy calibrations, and additional mitigation of backgrounds. 

\begin{acknowledgments}
\sortofsection{Acknowledgements}
We thank Tien-Tien Yu for helpful correspondence on the DM-electron scattering form factors. 
We gratefully acknowledge support from the National Science Foundation, Swiss National Science Foundation, German Ministry for Education and Research, Max Planck Gesellschaft, Deutsche Forschungsgemeinschaft, Netherlands Organisation for Scientific Research (NWO), Netherlands eScience Center (NLeSC) with the support of the SURF Cooperative, Weizmann Institute of Science, Israeli Centers Of Research Excellence (I-CORE), Pazy-Vatat, Fundacao para a Ciencia e a Tecnologia, Region des Pays de la Loire, Knut and Alice Wallenberg Foundation, Kavli Foundation, and Istituto Nazionale di Fisica Nucleare. 
This project has received funding or support from the European Union’s Horizon 2020 research and innovation programme under the Marie Sklodowska-Curie Grant Agreements No. 690575 and No. 674896, respectively.
Data processing is performed using infrastructures from the Open Science Grid and European Grid Initiative. We are grateful to Laboratori Nazionali del Gran Sasso for hosting and supporting the XENON project. 
\end{acknowledgments}

\bibliographystyle{apsrev4-1}

%

\clearpage

\onecolumngrid
\appendix*
\section{Supplemental Material}

\begin{table*}[h]
\centering
\begin{tabular}{@{}lllrllll@{}}
    \toprule
    Model                & $m_\chi$ [\si{GeV/c^2}]                       & $\sigma$ [\si{{cm}^2}]                    & \multicolumn{2}{c}{S2 ROI [PE]} & Observed & Expected signal  & Expected bg. \\ \colrule
    SI NR &  4 & \SI{6.4e-43}{} & {[}165,       & 271{]}       & 16    & $24.6 \pm 4.0$ & $3.7 \pm 0.6$ \\
    SI NR &  6 & \SI{4.8e-44}{} & {[}165,       & 276{]}       & 16    & $23.3 \pm 3.2$ & $4.0 \pm 0.6$ \\
    SI NR & 17 & \SI{1.8e-45}{} & {[}201,       & 578{]}       & 10    & $12.4 \pm 1.4$ & $5.9 \pm 0.8$ \\
    SI NR & 18 & \SI{5.2e-46}{} & {[}306,       & 1190{]}       & 3    & $4.4 \pm 0.5$ & $3.7 \pm 0.6$ \\
    SI-LM, $m_\phi = \SI{1}{MeV/c^2}$ & 4 & \SI{2.9e-38}{} & {[}165,       & 271{]}       & 16    & $24.8 \pm 4.2$ & $3.7 \pm 0.6$ \\
    SD (neutron-only) & 4 & \SI{9.8e-38}{} & {[}165,       & 271{]}       & 16    & $24.6 \pm 4.0$ & $3.7 \pm 0.6$ \\
    DM-electron scattering & 1 & $\SI{2.5e-40}{}$ & {[}165,       & 271{]}       & 16    & $24.7 \pm 4.1$ & $3.7 \pm 0.6$ \\ \colrule
    ALP & \SI{0.5e-6}{} & $g_\mathrm{ae} = \SI{7.7e-14}{}$ & {[}653,       & 897{]}       & 0    & $2.7 \pm 0.3$ & $0.5 \pm 0.1$ \\
    Dark photon & \SI{0.5e-6}{} & $\kappa = \SI{2.2e-16}{}$ & {[}653,       & 897{]}       & 0    & $2.7 \pm 0.3$ & $0.5 \pm 0.1$ \\
    \botrule 
    \\ 
\end{tabular}
\caption{Results for specific DM models excluded at exactly 90\% confidence level. The columns show successively: DM model (abbreviated as in the text, LM denotes light mediator and ALP axion-like particle); DM mass $m_\chi$; interaction strength, specified by the cross section for elastic scattering of DM with a free nucleon (or an electron, for DM-electron scattering) at zero momentum transfer for most models, by the axio-electric coupling constant $g_\mathrm{ae}$ for ALPs, and the kinetic mixing parameter $\kappa$ for dark photons; the left and right bounds of the S2 ROI in photoelectrons (PE); the number of observed events in the ROI; the expected number of events from the excluded signal model; the expected number of background events from the modeled backgrounds. The final two columns also show aggregated systematic uncertainties.}
\label{tbl:intervals}
\end{table*}

\twocolumngrid

\clearpage
$\!$
\newpage
$\!$
\pagebreak
$\!$

\subsection{Details on the energy scale}

\begin{figure}[h]
    \centering
    \includegraphics[width=0.49\textwidth]{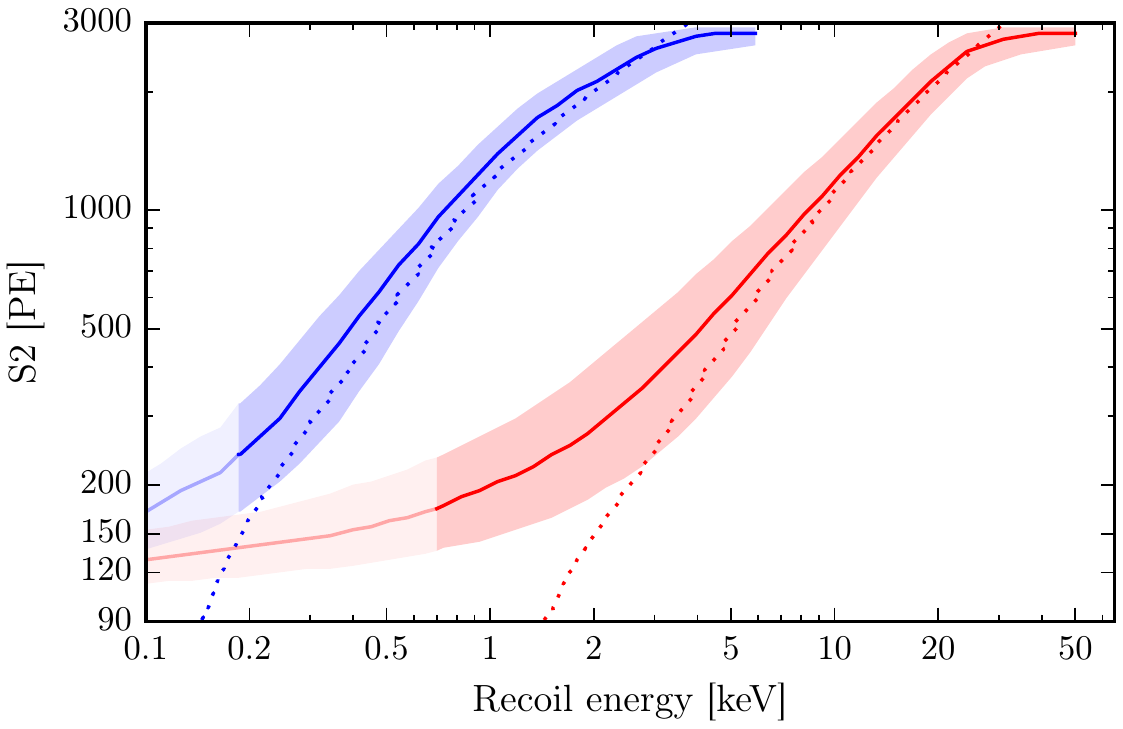}
    \caption{S2 response after all efficiencies for ER (blue) and NR (red). Solid lines show the median S2 response to signals of a specific energy, considering S2s in [90, 3000] PE only. The shaded regions indicate the $\pm 1 \sigma$ spread of the response. No NR (ER) response is assumed below 0.7 (0.186) \si{keV}; the hypothetical S2 response to those regions is shown lightly shaded for illustration. Dotted lines show the mean energy of events observed at a particular S2, in a flat (in linear energy) spectrum with no $Q_y$ cutoff.}
    \label{fig:s2_resolution}
\end{figure}

Figure \ref{fig:s2_resolution} shows the assumed response in the S2 region of this analysis to ER and NR signals at different energies and including all efficiencies. As stated in the text, for each dark matter model and mass, a specific S2 ROI is used, whose lower bound is never below 150 PE.

The average energy of events observed at a particular S2 depends on the underlying energy spectrum. For a flat energy spectrum, the majority of low-S2 events are underfluctuations, i.e.~events with an S2 below the median S2 response for their energy. Thus, the average energy of events from a flat spectrum observed at a given S2 is higher than a naive conversion based on the median S2 response of different energies would suggest. This is an instance of `Eddington bias', a systematic effect caused by statistical fluctuations. Similar effects are discussed in \cite{lux_dd}.

\vfill\eject

\subsection{Details on the S2 width model}

\begin{figure}[h]
    \centering
    \includegraphics[width=0.49\textwidth]{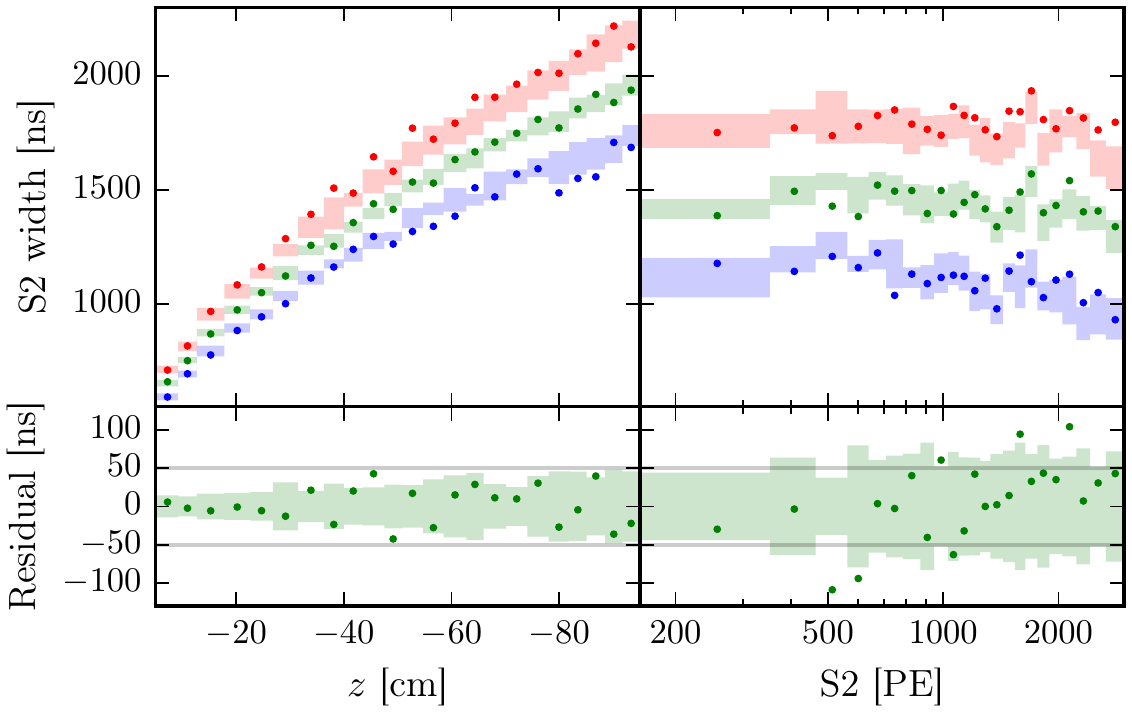}
    \caption{First quartile (blue), median (green) and third quartile (red) of S2 widths in simulation (bands) and neutron generator calibration data (dots), as a function of depth $z$ (left) and S2 size (right), integrated over the other dimension (S2 on the left, $z$ on the right). The lower panels show the same, for the median only, with the nominal model subtracted. This figure uses events with S1s (so their $z$ is known) that pass event selections besides those on $z$ and S2 width; the multiple-S2 and SE-before-S2 cuts were relaxed to obtain additional statistics. The width of the bands denotes the uncertainty due to limited calibration data statistics (100 events/bin).}
    \label{fig:width_validation}
\end{figure}

The S2 waveform width is defined as 
\begin{equation}
w = \big( \Delta t_{50\%} + \Delta t_r /2 \big) /2 ,
\end{equation}
where $\Delta t_{50\%}$ is the time between the first and third quartile of the S2 waveform, and $\Delta t_r$ the time between the start and median of the S2 waveform. This definition showed a slightly better correlation with depth than using $\Delta t_{50\%}$ alone.
We estimate the distribution of S2 widths using XENON1T's waveform simulator \cite{ap1}. Figure \ref{fig:width_validation} compares the S2 widths from simulation with those observed in neutron generator calibration data, showing that the median width agrees to within $\roughly\SI{50}{ns}$. S2 widths of \rnzero calibration data are in similarly good agreement. S1-tagged cathode events are broader than the model (by $\roughly \SI{120}{ns}$), as expected from the higher diffusion constant and lower drift velocity in the low-field region at and immediately above the cathode.

The decrease of the mean S2 width with S2 at a fixed depth, clearly visible in Figures~\ref{fig:monster} and \ref{fig:cathode} for cathode events, is not discernible in Figure~\ref{fig:width_validation} due to a compensatory selection effect. At low S2, events with S1s are more common deep in the detector, where the charge loss to impurities and the photon detection efficiency are largest, but the average S2 width is also larger.

\clearpage

\subsection{Details on the cathode background estimate}
Here we provide more details on our estimate of the background from decays on the cathode wires. As explained in the main text, most cathode events have detectable S1s even at very low S2. These cathode events with S1s are trivial to select, as they have a reconstructed depth $z$ within $\roughly \SI{1}{cm}$ of the cathode, which sits at a drift length of $\SI{-96.9}{cm}$, with a spread that results from imperfectly corrected effects of the drift field inhomogeneity. In contrast, these events are homogeneously distributed in $(x,y)$, and their rate shows no significant time-dependence during SR1. We also observe \ce{^{214}Po} and \ce{^{210}Po} alpha decays at the cathode, and thus hypothesize that $\beta$ decays from \ce{^{214}Pb} and especially \ce{^{210}Pb} (due to its low $Q = \SI{64}{keV}$) contribute to the observed low-energy cathode events.

\begin{figure}[b]
    \centering
    \includegraphics[width=\textwidth]{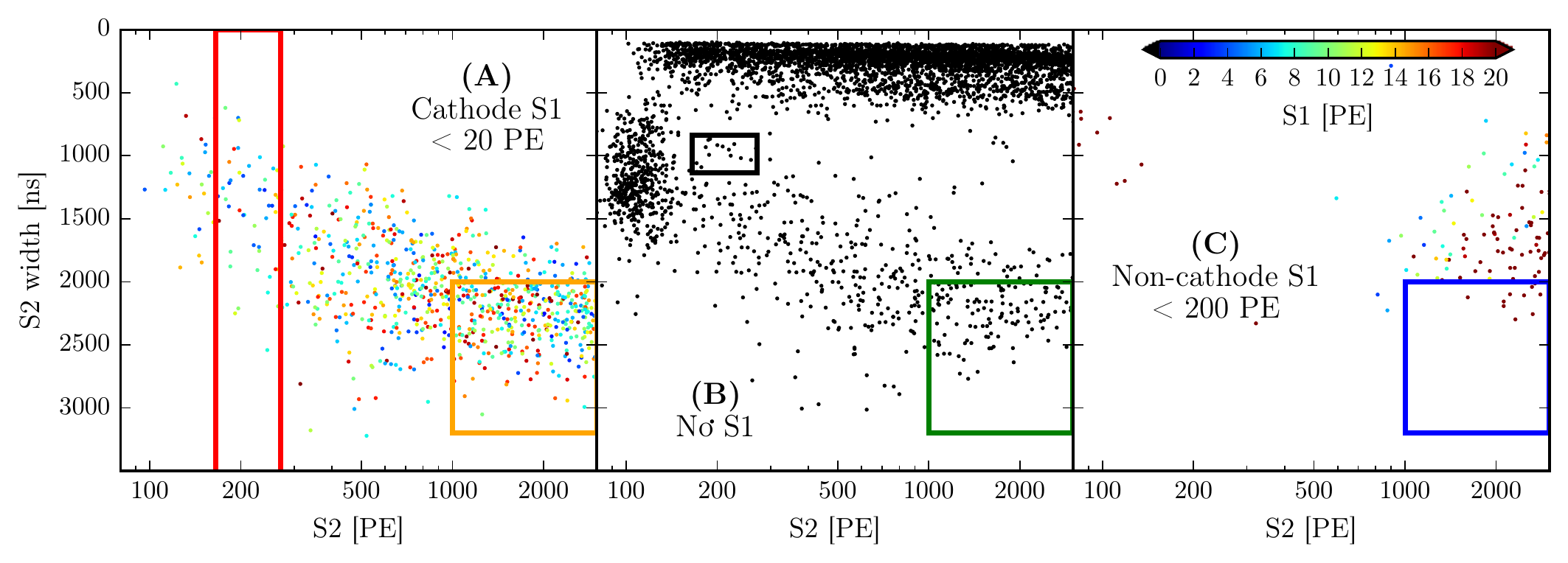}
    \caption{Illustration of the cathode background estimate. The scatterplots show (A) cathode events with $S1 < \SI{20}{PE}$, (B) events without S1s, and (C) events with $S1 < \SI{200}{PE}$ not from the cathode. Dot colors denote the S1 size in PE according to the colorbar in panel C. The colored boxes are used as described in the text to estimate the background in the DM ROIs after the width cut; the black box shows this region for \SI{4}{GeV/c^2} spin-independent NR DM. All events shown are from the search data, without the S2 width cut, and with the radial cut restricted to $R^2 < \SI{700}{cm}$ to avoid including misreconstructed events from the TPC walls in the blue and green boxes.}
    \label{fig:cathode}
\end{figure}

Figure~\ref{fig:cathode} illustrates the procedure for estimating the background from cathode events without S1s, after the width cut and inside the \SI{4}{GeV/c^2} NR analysis' S2 ROI, shown in \textbf{black} in figure \ref{fig:cathode}. First, we count cathode events with $S1 < \SI{20}{PE}$ that would appear in the S2 ROI if not for the width and z cuts. This region is shown in \textcolor{red}{red} in Figure~\ref{fig:cathode}. Next, we multiply this by the efficiency $\epsilon_w = (10.5 \pm 1.9) \%$ of the width cut on cathode events in the ROI, determined on cathode events with larger S1s ($20-\SI{200}{PE}$, not shown). Finally, we multiply by the ratio of cathode events without S1s and those with S1s $< \SI{20}{PE}$ in a control region at high S2 and high S2 width -- the count in the \textcolor{green}{green} divided by that in the \textcolor{orange}{orange} boxes in Figure~\ref{fig:cathode}. The count in the \textcolor{green}{green} box is corrected for (an insignificant) contamination by ordinary ER events. This is estimated by first counting events with non-cathode S1s in the control region -- the \textcolor{blue}{blue} box in Figure~\ref{fig:cathode} -- then scaling by one minus the S1 detection efficiency ($\epsilon_\mathrm{S1} > 90\%$) on ordinary ER events in the control region (determined as in \cite{ap1}). The cathode background estimate in the S2 ROI is thus:
\begin{equation} 
\textcolor{red}{r} \cdot \epsilon_w \cdot [\textcolor{green}{g} - \textcolor{blue}{b} \cdot (1 - \epsilon_\mathrm{S1}) ] / \textcolor{orange}{o} ,
\end{equation}
where $\textcolor{red}{r}$, $\textcolor{green}{g}$, $\textcolor{blue}{b}$ and $\textcolor{orange}{o}$ are the event counts in the red, green, blue and orange boxes, respectively. 

This procedure assumes that the ratio of cathode events with and without S1s is the same in the ROI as in the control region. However, events in the ROI have a lower S2, thus on average a lower energy, and are hence more likely to lack an S1 than events in the control region. Our estimate is therefore only a lower bound, and using it in the inference produces conservative limits.

We find $\textcolor{red}{r} = 46$, $\textcolor{orange}{o} = 336$, $\textcolor{green}{g} = 123$, $\textcolor{blue}{b} = 10$, leading to a cathode background estimate of $(1.62 \pm 0.46)$ events. The uncertainty includes statistical uncertainties on all the counts and the uncertainties on $\epsilon_w$ and $\epsilon_\mathrm{S1}$. The cathode background uncertainty is included in the aggregate background uncertainty described in the main text. We verified the estimate is stable well within its uncertainty under variations of the control region's S2 and S2 width bounds or the \SI{20}{PE} S1 threshold.

\onecolumngrid

\twocolumngrid

\end{document}